\documentclass[sigplan,screen]{acmart}
\title{Towards Formally Verified Compilation of Tag-Based Policy Enforcement}
\pdfoutput=1
\author{CHR Chhak}
\author{Andrew Tolmach}
\author{Sean Anderson}
\affiliation{%
\institution{Portland State University}
\city{Portland}
\state{OR}
\country{USA}
}
\email{
{chr.chhak,tolmach,ander28}@pdx.edu
}

\usepackage{breakurl}
\usepackage{hyperref}
\hypersetup{
    colorlinks,
    citecolor=black,
    linkcolor=black,
    urlcolor=blue,
    filecolor=black
}
\usepackage[makeroom]{cancel}
\usepackage{pgf, tikz}
\usepackage{graphicx}
\usepackage{verbatim}
\usepackage{xcolor, amsmath, mathtools, amsfonts}
\usepackage{relsize}
\usepackage{bussproofs}
\usepackage{multicol,stmaryrd}
\usepackage[section]{placeins}
\usepackage{ulem}
\usepackage{combelow}
\usepackage{microtype}
\usetikzlibrary{chains,shadows.blur,quotes}
\usetikzlibrary{matrix}
\usepackage{changepage}

\usepackage{xspace}

\newcommand{\blu}[1]{\shiftcolor{blue}{#1}}

\newcommand{\shiftcolor}[2]{\colorlet{current}{.}\color{#1} #2 \color{current}}

\newcommand{\tagine}{Tagine\xspace}
\newcommand{\pipe}{PIPE\xspace}

\newcommand{\compiler}{\ensuremath{\text{RTLgen}^{\text{T}}}\xspace}
\newcommand{\deadcodet}{\ensuremath{\text{Deadcode}^{\text{T}}}\xspace}
\newcommand{\cse}{\text{CSE}\xspace}
\newcommand{\cp}{\text{ConstProp}\xspace}

\newcommand{\source}{HLL\xspace}
\newcommand{\target}{\ensuremath{\text{RTL}^{\text{T}}}\xspace}

\newcommand{\itag}{I-tag\xspace}

\newcommand{\dfs}{\ensuremath{\mathcal{DFS}}\xspace}
\newcommand{\lpcp}{\ensuremath{\mathcal{PCP}}\xspace}
\newcommand{\wpci}{\ensuremath{\mathcal{WPCI}}\xspace}

\newcommand{\defl}{$\vcentcolon=$}
\newcommand{\deff}{\vcentcolon=}

\newcommand{\treval}{\Rightarrow}

\newcommand{\cminorsel}{CminorSel\xspace}
\newcommand{\rtlgen}{RTLgen\xspace}
\newcommand{\compcert}{CompCert\xspace}
\newcommand{\cakeml}{CakeML\xspace}

\newcommand{\eg}{e.g.,\xspace}

\newcommand{\ie}{i.e.,\xspace}
\newcommand{\etc}{etc.\xspace}

\newcommand{\op}{\oplus}

\newcommand{\ts}[1]{\mathit{ts_{#1}}}

\newcommand{\vat}{\textit{v@t}}

\newcommand{\pcc}{PC}
\newcommand{\pc}{\pcc\xspace}

\newcommand{\valtags}{\mathbf{T}}
\newcommand{\pctags}{\mathbf{P}}

\newcommand{\pct}{\textit{p}\xspace}
\newcommand{\pcp}{\textit{p}'}
\newcommand{\jpct}{\textit{p}_{\textit{s}}}

\newcommand{\oldt}{t_{\textit{old}}}

\renewcommand{\t}{\textit{t}}
\newcommand{\tp}{\textit{t}'}
\newcommand{\tone}{\textit{t}_1}
\newcommand{\ttwo}{\textit{t}_2}
\newcommand{\att}{{\textit{@t}}}
\newcommand{\attmore}[1]{{\textit{@t}#1}}
\newcommand{\attsub}[1]{{\textit{@t}_{#1}}}

\newcommand{\oxford}[1]{\llbracket #1 \rrbracket}

\newcommand{\updatedto}{\mapsto}
\newcommand{\cmp}{\bowtie}

\newcommand{\fsns}{fail-stop}
\newcommand{\fs}{\fsns\xspace}

\newcommand{\srule}[1]{\ensuremath{\overline{\texttt{#1}}}}
\newcommand{\oprule}{\ensuremath{\overline{\op}}}

\newcommand{\jrule}{\ensuremath{\overline{\textit{join}}}\xspace}

\newcommand{\wreturn}	{\texttt{Return}}

\newcommand{\wwhile}	{\texttt{While}}
\newcommand{\wif}	{\texttt{If}\ }
\newcommand{\wthen}	{\ \texttt{Then}\ }
\newcommand{\welse}	{\texttt{Else}\ }
\newcommand{\wskip}	{\texttt{Skip}}

\newcommand*{\defeq}{\mathrel{\vcenter{\baselineskip0.5ex \lineskiplimit0pt
                     \hbox{\scriptsize.}\hbox{\scriptsize.}}}%
                     =}

\newcommand{\wassign}{\ensuremath{\defeq}}

\newcommand{\wassignA}[2]{#1 \wassign #2}

\renewcommand{\v}{\textit{v}}

\newcommand{\wconsttr}	{\srule{Const}\xspace}
\newcommand{\wvartr}		{\srule{Var}\xspace}

\newcommand{\wassigntr}  	{\srule{\wassign}\xspace}

\newcommand{\wifsplittr}{\srule{IfSplit}\xspace}
\newcommand{\wifjointr}{\srule{IfJoin}\xspace}

\newcommand{\wwhileexittr}{\srule{WhileExit}\xspace}

\newcommand{\consttr}[3]{\srule{Const}\ #1\ #2 \treval #3}
\newcommand{\vartr}[3]{\srule{Var}\ #1\ #2 \treval #3}

\newcommand{\optr}[4]{\oprule\ #1\ #2\ #3\ \treval #4}
\newcommand{\asstr}[5]{\srule{\wassign}\ #1\ #2\ #3 \treval (#4,#5)}

\newcommand{\ifstr}[5]{\srule{IfSplit}\ #1\ #2\ #3\ #4 \treval #5}
\newcommand{\ifjtr}{\srule{IfJoin}}

\newcommand{\env}{E, \pct \vdash}
\newcommand{\enve}{E, \pct \vdash e}
\newcommand{\enveone}{E, \pct \vdash e_1}
\newcommand{\envetwo}{E, \pct \vdash e_2}
\newcommand{\envx}{E, \pct \vdash x}

\newcommand{\evalsto}{ \ \Downarrow \ }
\newcommand{\err}{{\textit{err}}}
\newcommand{\atom}{\v\att}
\newcommand{\atomprime}{\v\attmore{'}}
\newcommand{\atomone}{\v_1\attsub{1}}
\newcommand{\atomtwo}{\v_2\attsub{2}}

\newcommand{\emp}{\texttt{emp}}

\newcommand{\kjoin}[3]{ (#1, #2) ; #3}

\newcommand{\eval}[1]{{\fontfamily{Montserrat-TOsF}\selectfont
\scriptsize{$_{\textit{eval}}$\textit{#1}}}}

\newcommand{\evalconst}{\eval{Const}}
\newcommand{\evalvar}{\eval{Var}}
\newcommand{\evalop}{\eval{Op}}

\newcommand{\step}[1]{{\fontfamily{Montserrat-TOsF}\selectfont
\scriptsize{\textit{#1}}}}

\newcommand{\stepseq}			{\step{Seq}}
\newcommand{\stepskipseq}		{\step{SkipSeq}}
\newcommand{\stepskipjoin}	{\step{SkipJoin}}

\newcommand{\stepassign}			{\step{Assign}}

\newcommand{\st}{\mathcal{S}{\mskip-2mu}t}
\newcommand{\sregs}[6]{\mathcal{S}(#1,#2,#4,#3,#5,#6)}
\newcommand{\callstate}[4]{\mathcal{C}(#1,#2,#3,#4)}
\newcommand{\retstate}[3]{\mathcal{R}(#1,#2,#3)}
\newcommand{\errstate}{\mathcal{E}(\textit{err})}
\newcommand{\genv}{}
\newcommand{\cs}{c}

\newcommand{\trule}[1]{\ensuremath{\overline{\texttt{#1}}}}

\newcommand{\psregp}{pseudo-register}

\newcommand{\psregs}{\psregp s\xspace}

\newcommand{\tregs}[5]{\mathcal{S}(#1,#2,#3,#4,#5)}

\newcommand{\ropns}{\texttt{op}\ensuremath{_\oplus}}
\newcommand{\rnopns}{\texttt{nop}}
\newcommand{\rmovns}{\texttt{mov}}
\newcommand{\rmovins}{\texttt{movi}}
\newcommand{\rcondns}{\texttt{cond}\ensuremath{_{\cmp}}}
\newcommand{\rcallns}{\texttt{call}}
\newcommand{\rretns}{\texttt{ret}}

\newcommand{\rop}	{\ropns\xspace}
\newcommand{\rnop}	{\rnopns\xspace}
\newcommand{\rmov}	{\rmovns\xspace}
\newcommand{\rmovi}	{\rmovins\xspace}
\newcommand{\rcond}	{\rcondns\xspace}
\newcommand{\rcall}	{\rcallns\xspace}
\newcommand{\rret}	{\rretns\xspace}

\newcommand{\roptr}{\trule{\rop}\xspace}
\newcommand{\rnoptr}{\trule{\rnop}\xspace}
\newcommand{\rmovtr}{\trule{\rmov}\xspace}
\newcommand{\rmovitr}{\trule{\rmovi}\xspace}
\newcommand{\rcondtr}{\trule{\rcond}\xspace}

\newcommand{\nop}[1]{\texttt{nop}\ {#1}}
\newcommand{\mov}[3]{\texttt{mov}\ {#1}\ {#2}\ {#3}}
\newcommand{\movv}{\texttt{mov}\ r_s\ r_d\ n'}
\newcommand{\movi}[3]{\texttt{movi}\ {#1}\ {#2}\ {#3}}
\newcommand{\movii}[1]{\texttt{movi}\ {#1}\ r_d\ n'}
\newcommand{\opi}[4]{\texttt{op}_{\oplus}\ {#1}\ {#2}\ {#3}\ {#4}}
\newcommand{\opii}{\texttt{op}_{\oplus}\ \ r_1\ r_2\ r_d\ n'}

\newcommand{\condii}{\texttt{cond}_{\cmp}\ r_1\ r_2\ n_t\ n_f}

\newcommand{\gn}[1]{F_g(n) = {#1}\ \textit{@ itag} }
\newcommand{\dsp}[1]{{#1}\ \textit{itag}\ }
\newcommand{\Rb}{B}
\newcommand{\R}[3]{\Rb({#1}) = {#2} \textit{@} {#3}}

\newcommand{\exec}[1]{{\fontfamily{Montserrat-TOsF}\selectfont
\scriptsize{$_{\textit{exec}}$\textit{#1}}}}

\newcommand{\itij}{\texttt{ITifJoin}}
\newcommand{\itis}{\texttt{ITifSplit}}

\newcommand{\itspc}{\texttt{ITsavePC}}
\newcommand{\itass}{\texttt{ITassign}}

\newcommand{\itconst}{\texttt{ITconst}}
\newcommand{\itvar}{\texttt{ITvar}}
\newcommand{\itop}{\texttt{IT}}
\newcommand{\itdc}{\texttt{ITdc}}

\newcommand{\rpc}{r_{pc}}

\newcommand{\translex}[4]{n_{#1} \vcentcolon r_{#2} = \oxford{#3}_{n_{#4}}}

\newcommand{\instr}[3]{n_{#1}\vcentcolon #2 \ \textit{@}\ #3}

\newcommand{\translst}[3]{n_{#1} \vcentcolon \oxford{#2}_{n_{#3}}}

\newcommand{\ruletr}{\ensuremath{\overline{\textit{rule}}}}

\EnableBpAbbreviations
\setcopyright{licensedusgovmixed}
\acmPrice{15.00}
\acmDOI{10.1145/3437992.3439929}
\acmYear{2021}
\copyrightyear{2021}
\acmSubmissionID{poplws21cppmain-p32-p}
\acmISBN{978-1-4503-8299-1/21/01}
\acmConference[CPP '21]{Proceedings of the 10th ACM SIGPLAN International Conference on Certified Programs and Proofs}{January 18--19, 2021}{Virtual, Denmark}
\acmBooktitle{Proceedings of the 10th ACM SIGPLAN International Conference on Certified Programs and Proofs (CPP '21), January 18--19, 2021, Virtual, Denmark}



\begin{document}

\begin{abstract}
Hardware-assisted reference monitoring is receiving increasing attention as a way to improve the security of existing software. One example is the PIPE architecture extension, which attaches metadata tags to register and memory values and executes tag-based rules at each machine instruction to enforce a software-defined security policy.  To use PIPE effectively, engineers should be able to write security policies in terms of source-level concepts like functions, local variables, and structured control operators,  which are not visible at machine level.  It is the job of the compiler to generate PIPE-aware machine code that enforces these source-level policies.  The compiler thus becomes part of the monitored system’s trusted computing base---and hence a prime candidate for verification.

To formalize compiler correctness in this setting, we extend the source language semantics with its own form of user-specified tag-based monitoring, and show that the compiler preserves that monitoring behavior. The challenges of compilation include mapping source-level monitoring policies to instruction-level tag rules, preserving fail-stop behaviors, and satisfying the surprisingly complex preconditions for conventional optimizations. In this paper, we describe the design and verification of \tagine, a small prototype compiler that translates a simple tagged WHILE language to a tagged register transfer language and performs simple optimizations.
\tagine is based on the RTLgen and Deadcode phases of the \compcert compiler, and hence is written and verified in Coq.
This work is a first step toward verification of a full-scale compiler for a realistic tagged source language.
\end{abstract}

\maketitle

\section{Introduction} \label{intro}

\textit{Reference monitors}~\cite{refmon, refmon2} are a powerful mechanism for dynamic
enforcement of software security policies such as access control,
memory safety~\cite{refmonmem}, and information-flow control (IFC).
Monitors interpose a validation test at each security-relevant program
point and cause the program to fail-stop in the event of a security violation.
They are used in settings where the underlying software cannot easily be modified
or perhaps even inspected. This makes them an important tool of the
\textit{security engineer}---somebody tasked with improving system security,
often not the original programmer. However, monitoring is expensive to implement
in software, even when applied only at coarse granularity, e.g. only at 
function calls.

Recent work has shown that \textit{hardware-assisted} monitoring approaches can
enforce fine-grained security policies while still providing good performance.
For example, \pipe~\cite{PUMP-ASPLOS, stack} (Processor Interlocks for Policy Enforcement)\footnote{
  In previous and contemporaneous work, variants of \pipe are also called PUMP, SDMP, or CoreGuard.} is a
programmable hardware mechanism for supporting reference monitors at the granularity
of individual instructions. In a processor architecture extended with PIPE, metadata
tags are associated with each value in memory and registers. Just before each instruction
executes, PIPE checks its opcode and the tags on its operands to see if the operation
should be permitted, and if so, what tags should be assigned to the instruction's results.
These \textit{tag rules} collectively form a \textit{micro-policy}~\cite{micropol} (hereinafter
simply \textit{policy}). Tag rules are implemented in software (running in a privileged
supervisor context or on a dedicated co-processor), so policies are completely flexible
in how they interpret tags and gate machine operations.
Adding software checks at per-instruction granularity would be far too expensive, so the results of tag rules are stored in a fast hardware cache.
In a well-designed policy, the cache hit rate will be
high, so most instructions will execute at full speed. Experiments have shown reasonable
performance on a range of useful policies~\cite{PUMP-HASP, PUMP-ASPLOS,stack}.

However, the PIPE approach does have some limitations. Defining tag rules at the level
of the machine ISA is a difficult task for the security engineer, much as writing machine
code is harder than working in a high-level language. In principle, working at the
instruction level minimizes the trusted computing base (TCB) of the monitoring system;
in particular, security properties are enforced independently of how the machine code
was produced.
In practice, however, writing useful policies often requires understanding 
the output of a particular compiler. For example, a policy intended to guarantee
integrity of the stack~\cite{stack} must have at least partial knowledge of how the
compiler lays out stack frames and which generated instructions are performing
stack manipulation. This kind of reverse engineering is both tedious and error-prone.

More fundamentally, some policies can only be expressed in terms of high-level
code features that are not preserved at machine level. For example, an access control
policy might wish to gate entry to a function by inspecting the
tags on its arguments, but it may not be clear at machine level where those
arguments live. A memory safety policy may want to distinguish accesses to local variables
from accesses to the heap, even though both are compiled into the same machine-level load
and store instructions. Or an IFC policy may want to delimit the scope of implicit
flows~\cite{IFC-implicit} based on knowledge of the structured control flow (e.g. {\tt if-then-else}
constructs) in the source program, which is not explicitly visible in machine code.

We therefore propose defining policies in a high-level source language, compiling to
PIPE-compatible code, and including the compiler within the TCB. We extend a high-level
language with a tag-based reference monitoring semantics, and implement this extended
language by compilation to machine code for a PIPE-equipped processor. In the source
language, tag rules are triggered at meaningful \textit{control points} in the dynamic
semantics, such as evaluation of arithmetic operators, reading or writing variables,
function entry and return, and split and join points in the control-flow graph. 
We use hardware-level tags on the generated instructions to trace their provenance
back to the source-level construct (and associated control point and tag rule) that
produced them.

Since the compiler is now in the TCB, it is essential that it correctly
implements the intended monitoring semantics, in particular the fail-stop behavior.
So we verify it. In this paper, we present \tagine, a verified compiler that includes a
translator from a simple WHILE language
(with expressions, statements, and functions) to an instruction-level language of
control flow graphs, and a simple dead-code removal optimization for the instruction-level language.
\tagine is based on the RTLgen and Deadcode passes
of the \compcert C compiler~\cite{backend}; consequently, it is written in Gallina and
verified in Coq.  We have also implemented (though not verified) a tagged common-subexpression elimination (CSE)
optimization based on \compcert's \cse pass, and designed (though not implemented) a tagged version of \compcert's \cp pass.

Our initial work focuses on these compiler passes in order to study the most novel aspects of tagged compilation:
moving from source-level control points to per-instruction rules, and performing optimizations in the presence
of tag rules.
Our key verification result is \textit{policy preservation}: 
\tagine correctly preserves \fs behavior as well as standard semantics in the target code.
Although \tagine is currently lacking many important high-level language
features, notably memory and pointers, we believe it can be scaled up to a full compiler
for Tagged C, a version of C extended with control points and tagging that we are
currently designing.

This paper makes the following contributions:
\begin{itemize}
\item
  We describe a general scheme for implementing tag-based fine-grained reference monitoring in high-level language programs
  by compilation to \pipe-equipped hardware.
\item 
  We instantiate this scheme on simple source and target languages equipped with tag-based monitoring
  and implement the translation from source to target. 
\item 
  We verify in Coq that the translation preserves monitoring semantics.
\item
  We analyze the requirements for performing standard optimizations, including dead-code elimination,
  com\-mon-subexpression elimination, and constant propagation, in the tagged setting.
\item
  We implement and verify in Coq the dead-code elimination optimization, and implement the CSE optimization.
\end{itemize}
\noindent
The remainder of this paper is organized as follows.
\S \ref{pipe} gives background on the underlying \pipe tagged hardware architecture.
\S \ref{hll} shows how the idea of tagged monitoring can be extended to a high-level language.
\S \ref{compilation} outlines our general approach to compiling a tagged high-level language to \pipe.
\S \ref{prototype} formalizes \tagine's key pass, \compiler, and describes its verification. 
\S \ref{optimizations} discusses optimizations.
\S \ref{coqdev} gives a brief overview of our Coq development.
\S \ref{related} describes related work.
\S \ref{future} describes future work and concludes.
The complete Coq sources for \tagine may be found at \url{https://github.com/hope-pdx/Tagine-public}.

\section{\pipe}
\label{pipe}
\pipe is a collection of architectural features that extend a standard ISA (such as X86, ARM, or RISC-V) with support for tag-based,
per-instruction monitoring.  The design has been developed over the past eight years by a collaboration of industrial and academic 
researchers, partly under the aegis of several DARPA programs. Open-source hardware simulators and simple OS ports are 
available~\cite{hope-tools}, and IP incorporating the designs is currently marketed commercially by 
Draper Labs and Dover Microsystems~\cite{Dover20}.

\pipe augments architectural state by associating a \textit{metadata tag}
with each value in a register or memory location. Since instructions live in
memory, each instruction has a tag. In addition, the processor maintains a \textit{PC tag}
conceptually associated with the program counter value; this tag holds metadata characterizing
the current control state of the program.
Tags are intended to be large---roughly the size of pointers in the underlying architecture.
\pipe hardware makes no assumptions about the structure or meaning of tags,
which are completely configurable in software.

On each instruction, a \pipe-equipped processor evaluates a {\em tag rule} to determine
whether the instruction should be permitted to execute, and if so, what tags to put on
its result values. A distinct tag rule can be associated with each instruction op code; 
the inputs and outputs of the tag rule are op-code specific.
We write \srule{\texttt{op}} for the tag rule associated with instruction \ensuremath{\texttt{op}}.
For example, the RISC-V instruction \mbox{{\tt add} $r_d$,$rs_1$,$rs_2$}, 
which adds the contents of registers $\mathit{rs_1}$ and $\mathit{rs_2}$ and stores the result in register $\mathit{r_d}$, has a tag rule with signature
\[
\mbox{\tt \srule{add}:} (\mathit{ti},tpc,ts_1,ts_2) \rightarrow \mbox{\tt OK}(tpc',td) + \mbox{\tt Error}
\]
Here $\mathit{ti}$ is the tag on the {\tt ADD} instruction, $ts_1$ and $ts_2$ are the tags on source registers $rs_1$ and $rs_2$, and
$tpc$ is the current PC tag before the instruction executes.

The rule result is either {\tt OK} or {\tt Error}. In the {\tt Error} case,
the rule has decided that the instruction should not be permitted to execute,
and the processor halts or raises a software interrupt to terminate the process.
In the {\tt OK} case, execution continues, after setting two result tags:
$td$, the tag on the value written to destination register $rd$,
and $tpc'$, the new PC tag after the instruction executes.

As another example, the conditional branch instruction \mbox{{\tt beq} $rs_1$,$rs_2$,\mbox{\it offset}}
has the slightly simpler rule signature
\[
\mbox{\tt \srule{beq}:} (ti,tpc,ts_1,ts_2) \rightarrow \mbox{\tt OK}(tpc') + \mbox{\tt Error}
\]
because there is no result value to tag. The rule for the store instruction \mbox{{\tt stw} $rs_2$,\mbox{\it offset}($rs_1$)}
takes as an additional input the tag $tm$ of the old contents of the target memory location
and generates an additonal output tag $tm'$ for the new contents:
\[
\mbox{\tt \srule{stw}:} (ti,tpc,ts_1,ts_2,tm) \rightarrow \mbox{\tt OK}(tpc',tm') + \mbox{\tt Error}
\]
The tag rules for other instructions follow similar patterns.

A \textit{policy} is a complete collection of tag rules covering all the ISA's opcodes.
As a very simple example, we sketch an IFC policy intended to enforce confidentiality.
Suppose we wish to distinguish public and secret values and prevent the program
from writing secret values to certain memory locations $\mathcal{L}$ representing public channels.
To implement this scheme, we can use single boolean values
for both value and PC tags,
where {\tt true} means secret and {\tt false} means public.

We ignore instruction tags in this policy; their utility is explained in \S \ref{compilation}.
We assume that
values in memory have been pre-tagged appropriately; in particular, the
values in $\mathcal{L}$ are tagged {\tt false}.  New values computed from
secrets should also be secret. Also, to detect implicit flows, we maintain a ``security
context level'' in the PC tag; initially set to false, it is raised to true if we test
a secret value, since this can be used to expose the secret.  Here are some of the rules for this policy
(the other rules are similar):
\[
\begin{array}{ll}
  \mbox{\tt \srule{add}} (ti,tpc,ts_1,ts_2) \triangleq \\
  \mbox{\tt\ \ \ \ \ \ \ \ OK}(tpc'=tpc,td=ts_1 \lor ts_2 \lor tpc)\\
  \mbox{\tt \srule{beq}} (ti,tpc,ts_1,ts_2)  \triangleq \\
  \mbox{\tt\ \ \ \ \ \ \ \ OK}(tpc'=ts_1 \lor ts_2 \lor tpc)\\
  \mbox{\tt \srule{stw}} (ti,tpc,ts_1,ts_2,tm)  \triangleq \\
  \mbox{\tt\ \ \ \ \ \ \ \  if $(ts_2 \lor tpc) \rightarrow tm$} \\
  \mbox{\tt\ \ \ \ \ \ \ \  then OK}(tpc'=tpc,tm'=tm)\\
  \mbox{\tt\ \ \ \ \ \ \ \  else Error}\\
\end{array}
\]

One unfortunate feature of this policy is that once the PC tag has been raised by \srule{beq},
it remains secret indefinitely; this is a form of ``label creep''~\cite{sabelfeld03}.
While it would be sound to lower the
PC tag back to public when control reaches a join point following
both branches of the conditional, this is hard to do in a machine-level policy because
such join points are not explicit in machine code. We return to this issue
in \S \ref{hll}.

Software-defined policies are extremely flexible. The policy code can manage its
own data structures, even treating tags as pointers into its own (protected) memory space.
This is useful for combining policies by treating tags as (pointers to) data structures containing
the product of each policy's metadata. Policies can also maintain internal \textit{state} that persists
between rule invocations.
For example, a memory safety policy might maintain a counter to generate a fresh tag identifier for each object allocated in memory.

If every instruction of the \pipe-enhanced machine had to evaluate a tag rule in software before
executing, the system would be ridiculously slow. So \pipe relies on a
\textit{rule cache} which contains the results of recent rule evaluations, indexed by a tuple of
instruction opcode and input tags. The expectation is that in normal steady-state operation,
most instructions will find their tag rule result in the cache. The rule evaluation software
is invoked only in case of a cache miss. When designing policies, care must be taken
to avoid writing rules that inhibit effective caching.

\section{High-level language tag policies}
\label{hll}

We next consider how to lift the idea of tag-based policies from machine code to a 
higher-level language with  features such as expressions,
structured control flow, and functions. The key idea is to attach
tag rules to \textit{control points} in the language's execution semantics.
Control points are placed everywhere that a policy might want to inspect
tags and possibly halt execution. Tags themselves have arbitrary structure
and significance, just as at the PIPE hardware level, and we continue to assume
that rule evaluation is implemented in arbitrary software
(not necessarily coded in the high-level language being monitored).
The tag rule for a control
point is passed the tags of relevant values in the environment and,\blu{?} in some cases,
returns tags for result values. Also, even though there is no program counter
in a high-level language, we retain the idea of a ``PC tag'' that holds 
metadata associated with the current control state of the program;
it is passed to, and possibly updated by, each tag rule. 

For example, an assignment statement of the form \wassignA{$x$}{$e$} has a control point with a tag rule of the form
\[
\mbox{\tt \wassigntr:} (tpc,tx,te) \rightarrow \mbox{\tt OK}(tpc',tx') + \mbox{\tt Error}
\]
where $te$ is the tag on the value computed for $e$,  $tx$ and $tx'$ are
the tags on the contents of $x$ before and after the assignment, and 
$tpc$ and $tpc'$ are the PC tags before and after the assignment.
Note that this rule closely resembles the machine-level rule we saw above for {\tt STW},
which is not surprising given that an assignment might well be compiled into a store.

Similarly, each binary arithmetic expression \mbox{\tt $l$ $\op$ $r$}
has a control point that triggers a tag rule
\[
\mbox{\tt \oprule:} (tpc,tl,tr) \rightarrow \mbox{\tt OK}(tpc',t') + \mbox{\tt Error}
\]
where $tl$ and $tr$ are the tags of $l$ and $r$, $tpc$ and $tpc'$ are the
PC tags before and after evaluation, and $t'$ is the tag to be associated with the
result of the operation.

(For a language in which expression evaluation cannot change program state, it
might make sense to prevent expression tag rules from changing the PC tag,
in which case $tpc'$ would not be included as part of the rule result.)
This rule is similar to the machine-level {\tt ADD} rule, which again is unsurprising.

The control points for structured control statements are more novel. The basic
idea is to place a control point wherever the control flow graph splits or joins.
For example, an {\tt if-then-else} statement has two control points, one at the
conditional test point and another at the join point following the statement:

\begin{minipage}{3in}
\noindent
\\
\verb+      if+ $e_l \cmp e_r$ $\ \longleftarrow$ \wifsplittr\\
\verb+      then+ $s_1$\\
\verb+      else+ $s_2$\\ 
\verb+      endif+$\ \longleftarrow$  \wifjointr\\
\end{minipage}

\noindent
The associated tag rule forms are:
\[
\begin{array}{l}
\mbox{\tt \wifsplittr} (tpc,\cmp, t_l,t_r) \rightarrow \mbox{\tt OK}(tpc') + \mbox{\tt Error} \\
\mbox{\tt \wifjointr} (tpc,tpc_0) \rightarrow \mbox{\tt OK}(tpc') + \mbox{\tt Error} 
\end{array}
\]
where $\cmp$ is an arbitrary binary comparison, $t_l$ and $t_r$ are the tags of the compared values, $tpc$ and $tpc'$ are the PC tags before and after rule execution, 
and $tpc_0$ is the original PC tag \textit{at the split point} corresponding to the join point being executed.
To show the motivation for this rule signature, consider again the IFC secrecy policy from \S \ref{pipe},
this time expressed using high-level language tag rules.
\[
\begin{array}{ll}
\mbox{\tt \oprule} (tpc,tl,tr) & \triangleq \mbox{\tt OK}(tpc'=tpc,t'=tl \lor tr \lor tpc)\\
\mbox{\tt \wassigntr} (tpc,tx,te) & \triangleq \mbox{\tt if $(te \lor tpc) \rightarrow tx$ then} \\
  & \mbox{\tt\ \ \ \ OK}(tpc'=tpc,tx'=tx)\\
  & \mbox{\tt\ \ else Error}\\
\mbox{\tt \wifsplittr} (tpc,te) & \triangleq \mbox{\tt OK}(tpc'= te \lor tpc)\\
\mbox{\tt \wifjointr} (tpc,tpc_0) & \triangleq \mbox{\tt OK}(tpc' = tpc_0)\\
\end{array}
\]
Recall that we use the PC tag to track the ``security context level,'' which needs to be raised
to secret ({\tt true}) when we are executing conditionally under control of a secret,
in order to detect implicit flows.
A key benefit of using high-level language tag rules here is that the \wifjointr control point
rule can reset the PC tag to its original value when control leaves the {\tt if} statement, thus
potentially allowing subsequent statements to execute at lower secrecy. This rule is only
sound because, unlike the machine-level \pipe, the high-level language monitoring framework understands the semantics of
structured control flow operators. Other structured statements like {\tt while} and {\tt case} need similar control points.

Finally, control points are also placed before and after each function call site and at each function
entry and exit.  Rules executed at these points can inspect the tags on function parameter values as
well as on the function itself. Again, this is also information that would be difficult or impossible
to collect at machine level.

Note that the set of control points and tag rule signatures will typically be fixed once and for all when designing
monitoring for the high-level language. They should therefore be designed to be sufficiently general to implement any policy of interest.
Our control point design is based primarily on consideration of IFC, memory safety, and compartmentalization policies.
Of course, adequacy of the control point design cannot be absolutely guaranteed, as new kinds
of policies may be invented later. 

\section{Compilation approach}
\label{compilation}

Tag-based high-level language policies could be monitored in software, e.g. by
generating code to evaluate the rule functions and interleaving it with normal
execution code in the spirit of aspect-oriented weaving~\cite{KiczalesLMMLLI97}.
But given the density of control points, the overhead of this approach would probably be very high.
We instead opt to \textit{compile} the tagged high-level language to
machine code for a \pipe-equipped processor, in such a way that the reference
monitoring behavior of the source is preserved in the target.

Such a \textit{policy-preserving} compiler can be built by modifying a standard
compiler from (untagged) source to (untagged) target.  The task is simplified
by the fact that the structure and meaning of tags is largely the same at both levels, and
invocations of the source-level tag rule evaluation code can be embedded
directly in the target-level rules.

The main challenge is that the high-level monitor associates
tag rules with (language-dependent) control points, whereas the \pipe framework
associates them with each individual machine instruction.
While some control points, such as those at arithmetic operations,
correspond naturally to single instructions, others will correspond to multiple
instructions. Moreover, many different high-level features will compile
to instructions that use the same opcodes. For example, an {\tt add} instruction
in the target code might be implementing an explicit addition expression in
the source code, but it might equally well have been generated by the compiler
as part of array addressing or stack frame management. Clearly the opcode alone
is not sufficient to determine which source tag rule should be executed
at a given target instruction.

To solve this problem, we rely on the fact that \pipe associates a separate
tag $ti$ with each instruction in memory, and feeds it as one of the inputs to the
rule evaluated at each execution step. 
Instruction tags ({\itag}s) effectively let us design a customized instruction set
that \textit{refines} the hardware ISA by providing different variants of
some opcodes based on the instruction's semantic role in the policy being enforced.
Here, we use {\itag}s to specify \textit{provenance}, i.e. the source code
construct from which the instruction was generated.  The opcode's tag rule can dispatch
on the \itag to evaluate the relevant source tag rule, if any,
for each possible provenance.  For example, if an {\tt add} instruction 
is generated from an explicit {\tt +} expression, it might be tagged 
\itop+, whereas otherwise it might be tagged \itdc\ (for ``don't care'').
The tag rule for {\tt add} could then be:
\[
\begin{array}{l}
  \mbox{\tt \srule{add}} (ti,tpc,ts_1,ts_2) \triangleq \\
  \mbox{\tt \ \ match $ti$ with} \\
  \mbox{\tt \ \ \ \ IT+ $\Rightarrow$ \srule{$+$}($tpc$,$ts_1$,$ts_2$)} \\
  \mbox{\tt \ \ \ \ \itdc\ $\Rightarrow$ OK($tpc'=tpc,td=ts_1$)} \\
\end{array}
\]
where in the ``don't care'' case we arbitrarily choose to propagate the left operand's
tag to the result tag.

This ``piggybacking'' technique, in which we trigger the source control point
rule check by attaching it to the \pipe-rule for an instruction that
is already being generated, will work for most source language constructs.
But sometimes a tag-aware compiler must generate additional instructions into
the target code just to manage tags.  One example of this is the
{\tt if-then-else} statement.  As described in \S \ref{hll}, the PC
tag at the control split point must be saved so it can be passed to the
\wifjointr at the join point.
To do this, we can generate a target instruction at the split point (for example a {\tt mov} with a particular \itag) whose only purpose is to copy the PC tag into (the tag portion of) a register or onto the stack until it is needed at the join point.\footnote{
  An alternative approach would be for the policy to maintain a stack of split point PC tags
  in the current PC tag itself, avoiding the need to plumb them through the rules.
  Although this would be simpler than our scheme, its worst-case rule cache efficiency would be
  poorer.
}
  Similar dummy instructions
may be needed to track the PC tag at other structured control statements, or when marshalling the tags of
function arguments to feed them to a tag rule at a function entry control point, etc.

To formalize correctness of a tag-aware compiler, we start by defining semantics
for source and target languages that incorporate tag-based monitoring by construction.  Both semantics are parameterized by tag policies; the source tag rules are arbitrary, and the target tag rules embed the source rules. 
Tag policy violations lead to \fs states, which are distinct from stuck states
or other kinds of errors. Then a policy-preserving compiler is one that
preserves both ordinary computation behaviors and \fs behaviors. In particular,
the compiler must \textit{not} treat source policy errors as undefined behaviors
that can be refined into arbitrary valid executions in the target.

In principle, a policy-preserving compiler can be completely ignorant of the actual source
language policy, so that a single version of the generated code can be used to
enforce arbitrary source policies just by changing the tag rule evaluation code.
To achieve this, the source and target must invoke the same sequence of source rules, with the same arguments;
since the rules are arbitrary, any change in a rule invocation might change the \fs behavior of the overall rule sequence.
In addition to maximizing runtime flexibility, this approach also keeps the compiler and
its verification relatively simple.

However, maintaining this \textit{policy-independence} property may slow down target
code unnecessarily.  For example, the compilation scheme for saving and restoring
PC tags described above introduces extra instructions and adds extra pressure on register use, which
may degrade performance: if the policy being run doesn't actually make use of the PC tag,
adding this overhead is pointless.
More subtly, the need to preserve arbitrary tag rule
semantics inhibits the applicability of many simple code optimizations such as
dead code elimination, common subexpression elimination, or constant folding and propagation. 
For example, a conventional optimizer might use a standard liveness analysis to
eliminate an {\tt add} instruction if its result register value is never used.
However, in general it is not sound to skip evaluation of the instruction's
associated tag rule. Although calculating the tag on the instruction result
is not important---since that result is never used, its tag is not read either---the rule might fail-stop, change the PC tag, or change internal policy state.

In general, determining statically whether a tag rule execution can be skipped is
clearly uncomputable. 
However, our analysis of the existing CompCert optimizations on scalar locals has identified
several simple and intuitive conditions on tag rules that, in various combinations,
suffice to keep these optimizations sound.
These conditions might include not fail-stopping, not altering the PC tag, or being insensitive to the input PC
tag.  Thus, to enable optimizations, \tagine must know at least something about 
the rules, but not necessarily have their full definitions. Adopting this condition-based
approach helps decouple the compiler from the details of the rules, and will allow
the same compiled code to run with multiple sets of rules as long as they obey the conditions.

\section{The \compiler compiler pass}
\label{prototype}

To study the verification of policy preservation in detail, we have designed
and verified \compiler, a small prototype compiler pass that
translates \source, a simple tagged WHILE language (with expressions, statements, and functions) to \target,
an instruction-level language represented in an explicit control-flow graph (CFG). 
We focus on this compiler pass because it is here that high-level program structures
(statements and expressions) are transformed into instructions, and hence
where the tag rules for control points must be attached to appropriate instruction positions.

\source, \target and \compiler are closely based on the CminorSel and RTL languages and the \rtlgen pass of the \compcert compiler~\cite{backend},
and our proof of policy preservation is structured similarly to Leroy's correctness proof. 
We prove a forward simulation result which lifts into a refinement result thanks to the determinism of the target language~\cite{backend}.
As usual, this proof involves establishing and maintaining a matching relation between
corresponding source and target states.
We believe this is one of the more challenging parts of producing
a full-scale \compcert variant for Tagged C,  which is our long-term goal.

\subsection{The Source Language : \source} \label{source}

\source is a simple, untyped, deterministic, imperative language with expressions, structured statements, and functions with local variables.
For simplicity, we assume that the language has no I/O facilities, but the final value returned by the {\tt main} function
is observable.

\source's semantics is implicitly parameterized by a high-level rule policy $\mathcal{P}$,
which consists of a set of value tags $\valtags$, ranged over by $t$; 
a set of PC tags $\pctags$, ranged over by $\pct$;
a set of tag rules covering all possible control points;
and a set of of possible tag errors carried by \mbox{\tt ERROR} returns (which we elided for
simplicity in \S\ref{pipe} and \S\ref{hll}), ranged over by $\err$.

\begin{figure}
\begin{tabular}{@{}r@{\hspace{8pt}}l@{\hspace{2em}}l}
$e$ \defl	& \vat		& literal constants \\
$|$			& $x$		& local variables \\
$|$			& $e \op e$	& $\op \vcentcolon= + \ | \ -$ \\
\\
$s$ \defl	& \wskip							& empty/no-op\\
$|$			& \textit{s} ; \textit{s}				& sequencing\\
$|$			& $x \wassign e$								& assignment (to local)\\
$|$			& $\wif (e \cmp  e)\ \wthen s\ \welse s$	& ifs/conditionals\\
$|$			& $\wwhile\ (e \cmp e)\ s$				& while loops\\
$|$			& $x = f(\vec{e})$						& function calls\\
$|$			& \wreturn($e$)							& function returns\\
\\
$\cmp$ \defl & $\texttt{==} \ | \ \texttt{!=} \ | \ \texttt{<=} \ | \ \texttt{<} \ | \ \texttt{>=} \ | \ \texttt{>}$ & relational operator\\
\end{tabular}

\caption{Syntax of \source expressions and statements}
\label{source-syntax}
\end{figure}
literal constant expressions are
\textit{atoms}~\cite{PIPE-IFC}, consisting of a natural number $v$ paired with a value tag $t$, written $\vat$
and ranged over by $a$.
\source has explicit WHILE loops in place of \cminorsel's general-purpose
LOOP, BLOCK, and EXIT statements, because it is difficult to design sensible
IFC policies for the latter.
A function definition consists of a parameter list, 
local variable declarations,
body (a statement), 
and a function tag $p \in \pctags$ which will be made available to the tag rule
executed at function entry.
A program is just a collection of functions, with a distinguished {\tt main} function.

\begin{figure}[!tb]

  
\begin{prooftree}
\RightLabel{[\evalconst]}
\AXC{$\consttr{\pct}{\t}{\tp}$}
\UIC{$\env \atom \evalsto \atomprime$}
\end{prooftree}

\begin{prooftree}
\RightLabel{[\evalvar]}
\AXC{$E(x) = \atom$}
\AXC{$\vartr{\pct}{\t}{\tp}$}
\BIC{$\envx \evalsto \atomprime$}
\end{prooftree}

\begin{prooftree} \alwaysNoLine
\AXC{$\enveone \evalsto \atomone$}
\UIC{$\envetwo \evalsto \atomtwo$}
\AXC{$\optr{\pct}{\tone}{\ttwo}{\tp}$} \alwaysSingleLine \RightLabel{[\evalop]}
\BIC{$\enveone \op e_2 \evalsto (\v_1 \op \v_2) \attmore{'}$}
\end{prooftree}

    \caption{Evaluation of \source expressions (excluding errors)}
    \label{evalexpr}
\end{figure}

\begin{figure*}[th]
    \centering

\begin{prooftree}
\RightLabel{[\stepseq]}
\AXC{}
\UIC{$\genv \sregs{F}{(s_1;s_2)}{\pct}{k}{c}{E} \rightarrow
            \sregs{F}{s_1}{\pct}{(s_2;k)}{c}{E}$}
\end{prooftree}

\begin{prooftree}
\RightLabel{[\stepskipseq]}
\AXC{}
\UIC{$\genv \sregs{F}{\wskip}{\pct}{(s;k)}{c}{E} \rightarrow
            \sregs{F}{s}    {\pct}   {k} {c}{E}$}
\end{prooftree}

\begin{prooftree}
\RightLabel{[\stepskipjoin]}
\AXC{$\jrule\ \pct \ \jpct \treval \pcp$ }
\UIC{$\genv \sregs{F}{\wskip}{\pct}{\kjoin{\jrule}{ \jpct}{k}}{c}{E} \rightarrow
            \sregs{F}{\wskip}{\pcp}                              {k} {c}{E}$}
\end{prooftree}

\begin{prooftree}
\RightLabel{[\stepassign]}
\AXC{$\enve \evalsto \atom$ }
\AXC{$E(x) = \_\textit{@}\oldt$ }
\AXC{$\asstr{\pct}{\oldt}{t}{\pcp}{t'}$ }
\TIC{$\genv \sregs{F}{(x \wassign e)}{\pct}{k}{c}{E} \rightarrow
            \sregs{F}{\wskip}  {\pcp}{k}{c}{E[x \updatedto \atom']}$}
\end{prooftree}






\begin{prooftree}
\RightLabel{[\step{Cond}]}
\AXC{$\enveone \evalsto \atomone$}
\AXC{$\envetwo \evalsto \atomtwo$}
\AXC{$\ifstr{\pct}{\cmp}{t_1}{t_2}{\pcp}$}
\AXC{$s = \text{if\ } v_1 \cmp\ v_2 \text{\ then\ } s_t \text{\ else \ } s_f$}
\QuaternaryInfC{$\genv \sregs{F}{\wif (e_1 \cmp e_2)\ \wthen s_t\ \welse s_f}{\pct}                     {k} {c}{E} \rightarrow
                       \sregs{F}{s}                                         {\pcp}{\kjoin{\ifjtr}{\pct}{k}}{c}{E}$}
\end{prooftree}

    \caption{Selected \source statement transitions.}
    \label{sourcestepsstmt2}
\end{figure*}

Following \cminorsel, we give a relational natural semantics for expressions, and a transition system for statements and functions. 

\label{exprsem}
For expressions,  we define the judgement
\[\enve \evalsto (a + \err)\]
where $e$ is the expression being evaluated,
$E$ is the current environment, mapping variable to atoms,
and $\pct$ is the current \pc tag.
The result of evaluation is either an atom $a$ or a tag error $\err$.
Figure~\ref{evalexpr} gives the non-error cases for this judgement, which are standard
except for the tags and tag rule invocations. 
We write $\treval$ to indicate tag rule evaluation. Henceforth we use metavariables (without explicit \mbox{\tt OK} and \mbox{\tt ERROR} constructors) to indicate the type of rule results, and adopt a juxtaposition-as-application style for rule arguments.

Since expressions are pure,
we make the assumption that no policy will ever need
expression evaluation to change the \pc tag.
We omit the error cases induced when a tag rules returns an \err; as usual,
propagation of errors leads to an explosion of judgements. The error rules 
fix the order of sub-expression evaluation, since an error in the left
operand is propagated in favor of one in the right operand.

The semantics for \source statements is given by a transition system
between \textit{program states} $\st$.
The transition relation, written $\st \rightarrow \st'$, describes a single execution step.
Borrowing from \compcert, we distinguish function internal, entry, and exit states, and add a new state $\errstate$ representing a {\fs} due to tag error $\err$.\\

\begin{tabular}{@{}r@{\hspace{4pt}}l@{\hspace{4em}}l}
$\st$ \defl		& $\sregs{F}{s}{\pct}{k}{c}{E}$	& regular states	\\
$|$            	& $\callstate{F}{\vec{a}}{p}{c}$	& call states	\\
$|$            	& $\retstate{a}{p}{c}$			& return states	\\
$|$            	& $\errstate$					& \fs states		\\
\\
\end{tabular}

\noindent
Regular states correspond to execution within a function and carry
the current function $F$;
the current program point, represented by a statement-under-focus $s$ and a local continuation $k$;
the current PC tag $\pct$;
a call continuation $c$, representing the call stack;
and a local environment $E$ that maps variables to atoms.
In call states, $F$ is the callee, and $\vec{a}$ its parameters.
Return states carry a returned atom, $a$.

\compcert, following Appel and Blazy~\cite{appel2007separation}, combines local and callstack continuations into one continuation, but we find it simpler and clearer
to separate them.
Local continuations obey the following grammar:\\

\begin{tabular}{@{}l@{\hspace{6pt}}r@{\hspace{4pt}}l@{\hspace{2em}}l}
$k$	& \defl	& \emp								& return \\
	& $|$	& $s;\ k$							& continue with \textit{s}, then do \textit{k} \\
	& $|$	& $\kjoin {\jrule} {\jpct}{k}$	& update PC tag; then do $k$\\
	& & &\\
\end{tabular}

\noindent
The novelty here is the PC tag update, which is explained below.
To save space and remain focused on the key ideas of tag-based compilation, 
we will not discuss function calls, function entry and exit states, or call continuations further in this paper;
for details, see the full Coq development.

An \textit{initial state} (for a program) is the call state where $F$ is \texttt{main()} (which takes no parameters) and $c$ is empty.
Program execution is described by the transitive closure of the steps taken in the semantics from the initial state.
\textit{Final states} are return states with an empty call stack, and are those in which a program is considered to have terminated normally.
There are no transitions out of \fs states or final states.

A program may exhibit one of the following \textit{behaviors}:

\begin{itemize}
\item
	\textbf{Terminate} with result $a$, when program execution reaches a final state carrying $a$.

\item
	\textbf{Fail-stop} with tag error $err$,  when program execution reaches an error state carrying $err$.

\item
	\textbf{Diverge}, when program execution may always take another step in the transition semantics.

\item
	\textbf{Go Wrong} (or, ``get stuck''), when program execution cannot take a step in the transition semantics (but is not in a \fs, or final state).
\end{itemize}
\noindent
We write $S_{\mathcal{P}} \Downarrow B$ to mean that a program $S$, executing under policy $\mathcal{P}$, exhibits behavior $B$.
We use behaviors to help formalize a notion of semantic preservation (\S\ref{vprelim}).

Figure~\ref{sourcestepsstmt2} gives transition judgements for a small selection of statements.
The non-tag aspects of these are standard. 
Note that the local continuation grows under sequencing,
and is consumed when the statement under focus is \wskip.
The semantics precisely specifies the position
and signature of each control point. For example, \stepassign\ evaluates the
right-hand side to an atom with tag $t$,
fetches the tag $\oldt$ of the existing value
in $x$, and then passes them to \wassigntr together with the
PC tag $\pct$.  If the rule does not fail, it returns the new PC tag $p'$ and
a tag $t'$ to associate with the new value in $x$; we step to a new
regular state with the new PC tag and the environment entry for $x$ suitably updated.
\step{AssignRuleErr}\ (not shown) applies when 
\wassigntr returns a tag error $\err$, in which case we 
step to $\errstate$.

\begin{figure*}[!tbh]
    \centering

\begin{prooftree}
\RightLabel{[\exec{Nop}]}
\AXC{$\gn{\nop{n'}}$}
\AXC{$\dsp{\rnoptr} \pct \treval \pcp$}
\BIC{$\genv \tregs{F}{n} {\pct}{\cs}{\Rb} \rightarrow
            \tregs{F}{n'}{\pcp}{\cs}{\Rb}$}
\end{prooftree}

\begin{prooftree}
\alwaysNoLine
\AXC{$\gn{\movv}$}
\AXC{$\R{r_s}{\v_s}{t_s}$}
\UIC{$\R{r_d}{\v_d}{t_d}$}
\AXC{$\dsp{\rmovtr} \pct\ \t_s\ \t_d \treval (\pcp, \tp)$}
\alwaysSingleLine
\RightLabel{[\exec{Mov}]}
\TIC{$\genv \tregs{F}{n} {\pct}{\cs}{\Rb} \rightarrow
            \tregs{F}{n'}{\pcp}{\cs}{\Rb[r_d \updatedto \v_s \attmore{'}]}$}
\end{prooftree}



\begin{prooftree}
\RightLabel{[\exec{MovI}]}
\AXC{$\gn{\movii{\atom}}$}
\AXC{$\dsp{\rmovitr} \pct\ \t \treval (\pcp, \tp)$}
\BIC{$\genv \tregs{F}{n} {\pct}{\cs}{\Rb} \rightarrow
            \tregs{F}{n'}{\pcp}{\cs}{\Rb[r_d \updatedto \atomprime]}$}
\end{prooftree}

\begin{prooftree}
\alwaysNoLine
\AXC{$\gn{\opii}$}
\AXC{$\R{r_1}{\v_1}{t_1}$}
\AXC{$\R{r_2}{\v_2}{t_2}$}
\AXC{$\dsp{\roptr} \pct\ t_1\ t_2 \treval (\pcp,\tp)$}
\alwaysSingleLine
\RightLabel{[\exec{Op}]}
\QuaternaryInfC{$\genv
\tregs{F}{n} {\pct}{\cs}{\Rb} \rightarrow
\tregs{F}{n'}{\pcp}{\cs}{\Rb[r_d \updatedto (\v_1 \op \v_2) \attmore{'}]}$}
\end{prooftree}


\begin{prooftree}
\alwaysNoLine
\AXC{$\gn{\condii}$}
\AXC{$\R{r_1}{\v_1}{t_1}$}
\UIC{$\R{r_2}{\v_2}{t_2}$}
\AXC{$\dsp{\rcondtr} \pct\ \tone\ \ttwo \treval \pcp$}
\UIC{$n' = \text{if\ } \v_1 \cmp\ \v_2 \text{\ then\ } n_t \text{\ else \ } n_f$}
\alwaysSingleLine
\RightLabel{[\exec{Cond}]}
\TIC{$\genv
\tregs{F}{n}  {\pct}{\cs}{\Rb} \rightarrow
\tregs{F}{n'}{\pcp}{\cs}{\Rb}$}
\end{prooftree}





    
    \caption{Selected \target instruction transitions}
    \label{targetsteps1}
\end{figure*}

\begin{figure}

\begin{tabular}{@{}l@{\hspace{5pt}}r@{\hspace{2pt}}l@{\hspace{20pt}}l}
& \textit{i} $\vcentcolon=$ & $\rnop\ \textit{n}$                                  & no-op (go to \textit{n})\\
              &                       $|$ & $\rop\ r_{s1}\ r_{s2}\ r_d\ n$     & binary operation\\
              &                       $|$ & $\rmov\ r_s\ r_d\ n$                              & move\\
              &                       $|$ & $\rmovi\ \textit{a}\ \textit{r}_d\ \textit{n}$     & move immediate\\
              &                       $|$ & $\rcond\ r_1\ r_2\ n_t\ n_f$                  & branch\\
              &                       $|$ & $\rcall\ f\ \vec{r_s}\ r_d\ n$         & call\\
              &                       $|$ & $\rret\ \textit{r}$                               & return\\
 \end{tabular}\\

\caption{Syntax of \target instructions} 
\label{target-syntax}
\end{figure}

The control point semantics for {\tt if-then-else} are more complicated.
As discussed in \S \ref{hll}, we want one control point where the conditional
is evaluated and another at the implicit join point following the statement.
The first of these is specified by the invocation of \wifsplittr in the premises
of \step{Cond}. But the second join point needs to be
associated with the \textit{continuation} of the statement, and it needs to
be given the PC tag from the split point as one of its arguments.
To do this, we generate a continuation of the form $\kjoin{\jrule}{\jpct}{k}$.
(\jrule is a metavariable.)
This continuation indicates that the program passed through a split point earlier (in this case a conditional) and has now reached the corresponding join point.
It is processed by judgement \step{SkipJoin}, which invokes
$\jrule\ \pct\ \jpct$ ($\pct$ and $\jpct$ are the current and split-point PC tags, respectively),
updates the PC tag to this result, and proceeds with continuation $k$.
For the {\tt if-then-else} join we specify $\jrule$ to be \wifjointr
and $\jpct$ to be the split-point PC tag.
The same continuation mechanism is used for {\tt while} statements
(specifying, say, \wwhileexittr for $\jrule$).
A similar technique is used to specify the control points associated with
calls and returns.

\subsection{The Target Language: \target} \label{target}

\target is a deterministic, 3-address code, register transfer abstract machine language based on CFG's:
it represents functions as graphs, where each node is an individual instruction.
Like HLL, its semantics are parameterized by an instruction-level rule policy, i.e.
sets of value tags $\valtags$ and PC tags $\pctags$,
a set of tag rules covering each possible instruction, and
a set of possible error tags ranged over by $\err$.
In addition, \target is parameterized by an
arbitrary set of {\itag}s, ranged over by $\mathit{itag}$;
each instruction is labeled with an \itag.

Figure~\ref{target-syntax} describes the syntax of \target instructions. 
We use $n$ to range over labels for graph nodes and $r$ to range over \psregs.
Each instruction carries the label of the next node(s) to be executed.
We reuse the arithmetic and relational operators found in \source.
\noindent

An \target function $F$ is described by a graph $F_g$, which is a finite partial mapping from nodes
to instructions; an entrypoint node; registers containing parameters; and a function tag.
Each function has an infinite bank of registers.

Program states and behaviors are similar to those in \source.
Figure \ref{targetsteps1} gives transition judgements for \target instructions, omitting calls, returns, and error rules. 
Each instruction invokes a tag rule, which is passed the instruction's \itag in addition to value and PC tags.

\definecolor{orange}{RGB}{243, 181, 74}
\newcommand{\inlinenode}[1]{\tikz[baseline=(X.base)]\node [draw=black,fill=orange,rectangle,inner sep=1pt] (X) {$#1$};}
\newcommand{\tfence}[1]{$\mathlarger{\Downarrow}$\vspace*{6pt}\\}

\subsection{Compilation} \label{compiler}

We now describe compilation of \source into \target by example.
Compilation involves the translation of statements and expressions into instructions,
but also the injection of an \source tag policy $\mathcal{P}$ into an equivalent \target policy,  $\mathcal{I}(\mathcal{P})$.
The per-opcode rules in $\mathcal{I}(\mathcal{P})$ begin by
dispatching on {\itag}s: an instruction's tag rule effectively depends on both its opcode and its \itag.
For compactness, we write rules in the form
{\it opcode \itag parameters $\triangleq$ rule-body}.
Many {\itag}s correspond directly to source constructs, and their rule bodies simply invoke the corresponding \source tag rule.
Other {\itag}s signal administrative tag operations generated by the compiler.
We give some of the rules for $\mathcal{I}(\mathcal{P})$ below, interleaved with the discussion
of the relevant compilation cases.

The translation from \source to \target decomposes expressions into linear sequences of \target instructions,
and recursively translates statements into CFG's.
Functions and programs in \target are extremely similar to those in \source, so the compiler has little to do for these,
and we omit further discussion of them.

Since most \target instructions incorporate an explicit \textit{successor node}, the compiler builds CFG's in reverse execution order.
Each translation function takes a source fragment, a variable map (holding registers for parameters and locals), and a target successor node; it returns an entry node $n_e$ into the modified CFG.

{\bf Translation of expressions} 
We write the expression translation function as 
\[
\translex{e}{}{e}{s}
\]
where
$e$ is the expression being translated,
$n_s$ is the successor node,
$n_e$ is the entry node returned, and
$r$ is a fresh register generated to hold the result of $e$.
The output of translation is a CFG, rooted at $n_e$ and exiting to $n_s$, whose execution will have the effect of evaluating $e$
and (if this is successful) placing its atom in register $r$ before continuing to $n_s$.
If evaluation of $e$ leads to a tag error $\err$, execution of the subgraph will halt in state $\errstate$.

The translation function is defined by cases on syntax constructors.
We show the result of each translation as a CFG diagram.
Rounded white boxes represent single graph nodes showing an instruction, its \itag and node label.
Recursive calls to translation functions
generate subgraphs, represented by shaded rectangular boxes.

Here are the cases for constants, variables and operations,
where $r_x$ is the register mapped to hold the \source variable $x$:
\setlength{\columnsep}{1pt}
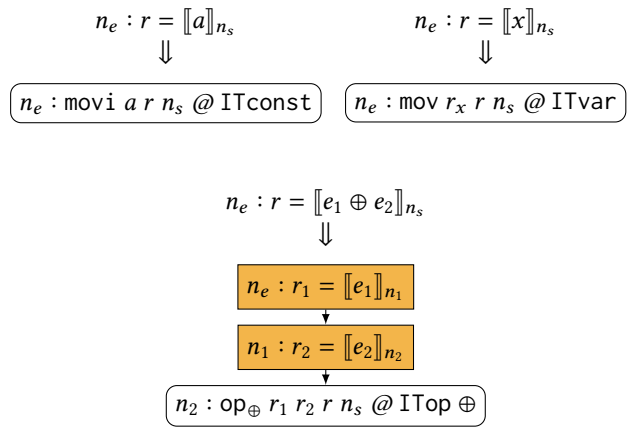
\begin{multicols}{2}
\centering
	$\translex{e}{}{a}{s}$\\
	\tfence{Constant}
    \begin{tikzpicture}[auto, box/.style = {draw,rounded corners,fill=white, align=center}]
        \node[box] (ne)        {$\instr{e}{\movi{a}{r}{n_s}}{\itconst}$};
        \label{compConst}
    \end{tikzpicture}
    
\columnbreak
\centering
    $\translex{e}{}{x}{s}$\\
    \tfence{Variable}

    \begin{tikzpicture}[auto, box/.style = {draw,rounded corners,fill=white, align=center}]
        \node[box] (ne)        {$\instr{e}{\mov{r_x}{r}{n_s}}{\itvar}$};
        \label{compVar}
    \end{tikzpicture}

\end{multicols}

\begin{center}

	$\translex{e}{}{e_1 \op e_2}{s}$\\
	\tfence{binary op}
    \begin{tikzpicture}[auto,
        node distance = 2mm,
        start chain = going below,
        box/.style = {draw,rounded corners,fill=white,
            on chain,align=center},
        rec/.style = {draw,fill=orange,
            on chain,align=center}]
    
    \node[rec] (ne)        {$      \translex{e}{1}{e_1}{1}$};
    \node[rec] (n1)        {$      \translex{1}{2}{e_2}{2}$};
    \node[box] (n2)        {$\instr{2}{\opi{r_1}{r_2}{r}{n_s}}{\texttt{ITop}\ \op}$};

    \begin{scope}[rounded corners,-latex]
    \path     (ne) edge (n1)      (n1) edge (n2);
    \end{scope}
    \label{compOp}
    \end{tikzpicture}

\end{center}

\noindent
To understand the tag-related behavior of the generated instructions, we must also examine the behavior of the
\target tag policy $\mathcal{I}(\mathcal{P})$, under which this code will execute.
These rules essentially replicate \source's tag processing in \target. The definitions for expressions are:\\
\begin{tabular}{lll@{\hspace{0.5em}}l}
  \\
\rmovitr		&$\itconst$& $p$ $t$ & $\triangleq$ \wconsttr $p$ $t$\\
\rmovtr		&$\itvar$ & $p$ $t_s$ $t_d$ &$\triangleq$ \wvartr $p$ $t_s$\\
\roptr		&$\itop \op$ &$p$ $ts_1$ $ts_2$ & $\triangleq
\begin{cases}
(\pct,\t)	& \text{if } \oprule\ \pct\ \ts{1}\ \ts{2} \treval\ \t\\
\err			& \text{if } \oprule\ \pct\ \ts{1}\ \ts{2} \treval\ \err
\end{cases}$

\end{tabular}
\\

\textbf{Translation of statements and functions}
The statement translation function is written
\begin{centering}
$\translst{e}{s}{s}$
\end{centering}
where $s$ is a source statement, $n_s$ a successor node parameter and $n_e$ the generated entry node.
Executing the resulting CFG, rooted at $n_e$ and exiting to $n_s$, will have the same effect in \target as statement $s$ does in \source.
As with expressions, if evaluation of $s$ leads to a tag error, execution halts in $\errstate$.
Assignment and sequencing CFG's and their rule definitions in $\mathcal{I}(\mathcal{P})$:

\vspace{0.2in}
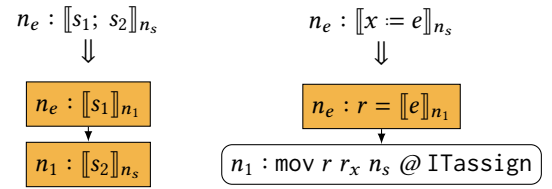
\begin{minipage}{1.3in}
\centering

	$\translst{e}{s_1;\ s_2}{s}$\\
	\tfence{Sequence}
    \begin{tikzpicture}[auto,
        node distance = 2mm,
        start chain = going below,
        rec/.style = {draw,fill=orange,
            on chain,align=center}]

   \node[rec] (ne)        {$ \translst{e}{s_1}{1}$};
   \node[rec] (n1)        {$ \translst{1}{s_2}{s}$};
 
   \begin{scope}[rounded corners,-latex]
   \path (ne) edge (n1);
   \end{scope}
   \label{compSeq}
 \end{tikzpicture}

\end{minipage}
\begin{minipage}{1.7in}
\centering

    $\translst{e}{x\ \text{\wassign}\ e}{s}$\\
	\tfence{Assignment}
    \begin{tikzpicture}[auto,
        node distance = 2mm,
        start chain = going below,
        box/.style = {draw,rounded corners,fill=white,
            on chain,align=center},
        rec/.style = {draw,fill=orange,
            on chain,align=center}]

    \node[rec] (ne)        {$      \translex{e}{}{e}{1}$};
    \node[box] (n1)        {$\instr{1}{\mov{r}{r_x}{n_s}}{\itass}$};
 
    \begin{scope}[rounded corners,-latex]
    \path (ne) edge (n1);
    \end{scope}
    \label{compAssign}
    \end{tikzpicture}

\end{minipage}
\vspace*{0.2in}

\begin{tabular}{llll}\\
\rmovtr &$\itass$& $p$ $t_s$ $t_d$ & $\triangleq$ \wassigntr $p$ $t_d$ $t_s$\\
\\
\end{tabular}

\noindent
Note that assignments and variable expressions both compile to a \rmov,
but the {\itag}s encode enough information about the source provenance of the instruction
to reproduce the correct rule processing in the target.

\textbf{Pseudo-instructions and join points}
The most interesting cases for statement compilation in our tagged world are conditionals and while loops; for brevity, we focus on the former.
Recall that these statements have multiple control points, corresponding to splits and joins in program control

\begin{center}

    $\translst{e}{\wif (e_1 \cmp e_2) \wthen s_t\ \welse s_f}{s}$\\
	\tfence{Conditional}
    \begin{tikzpicture}[auto,
    	node distance = 2mm,
    	box/.style = {draw,rounded corners,fill=white,
    		align=center},
    	rec/.style = {draw,fill=orange,
    		align=center}]
        
	\node[box] (ne)								{$\instr{e}{\mov{\rpc}{\rpc}{n_1}}{\itspc}$};
	\node[rec] (n1)	[below=of ne]				{$      \translex{1}{1}{e_1}{2}$};
	\node[rec] (n2)	[below=of n1]				{$      \translex{2}{2}{e_2}{3}$};
	\node[box] (n3)	[below=of n2]				{$\instr{3}{ \condii}{\itis}$};
	\node[rec] (nt)	[below=of n3, xshift=-7.5mm]	{$      \translst{t}{s_t}{4}$};
	\node[rec] (nf)	[right=of nt]				{$      \translst{f}{s_f}{4}$};
	\node[box] (n4)	[below=of nt, xshift=7.5mm]	{$\instr{4}{\mov{\rpc}{\rpc}{n_s}}{\itij}$};

	\begin{scope}[rounded corners,-latex]
	\path
	(ne) edge (n1)
	(n1) edge (n2)
	(n2) edge (n3)
	(n3) edge (nt.90)
	(n3) edge (nf.90)
	(nf.270) edge (n4)
	(nt.270) edge (n4);
	\end{scope}
	\label{compIf}
	\end{tikzpicture}

\end{center}

\begin{tabular}{llll}\\
\rmovtr	&$\itspc$ & $p$ \_ \_  &$\triangleq$ ($p$,$p$)\\
\rcondtr	&$\itis$ & $p$ $\bowtie$ $t_1$ $t_2$ & $\triangleq$ \wifsplittr $p$ $\bowtie$ $t_1$ $t_2$\\
\rmovtr	&$\itij$& $p$ $t_s$ $t_d$ & $\triangleq$ \wifjointr $p$ $t_s$\\
\\
\end{tabular}

\noindent
flow, and that the \pc tag at the split point needs to be passed as a parameter to the rule at the matching join point.
The compiled versions of these statements use pseudo-instructions to save the split point \pc tag (and a dummy value) and recover this \pc tag to use in join point rules.
These instructions are implemented as {\rmov}s and distinguished by their {\itag}s.

When used to save \pc tags at split points, \rmov ignores the tag of the source register, and moves the current \pc tag into the tag portion of the destination register.
When used to recover the \pc tag, it just passes the tag portion of the register ($t_s$) to the join point rule .
The pseudo-instructions for split and join points are always generated in pairs, and for each instruction, the same ``save'' register $\rpc$ is used for both source and destination, leaving the value part of $\rpc$ unchanged.
Since split-join pairs can be arbitrarily nested, the set of ``save'' registers that are live at any given program point form a stack.

\subsection{Verification Approach} \label{vprelim}
The verification of \compiler and \deadcodet (\S\ref{deadcode})  follow the general framework laid out in CompCert~\cite{backend}.
Our notion of semantics preservation is \textit{refinement-for-safe-programs}: Any behavior exhibited by the target program must be one exhibited by the source program. In other words, \textit{compilation should not introduce new behaviors in the target}. Behavioral equality is defined up to equality of results and errors, i.e., \textbf{Terminate} $a$ = \textbf{Terminate} $a'$ iff $a=a'$ and \textbf{Fail-stop} $err$ = \textbf{Fail-stop} $err'$ iff $err=err'$.
Further, we only want to consider \textit{safe} source programs, i.e. those that do not \textbf{Go Wrong}. This is to allow the compiler the flexibility of, for example, removing a division by zero whose result is unused.

Each \tagine pass is proved via a \textit{forward} simulation \\
\newcommand{\othersigma}{\varsigma}

\begin{center}
\begin{tikzpicture}
  \matrix (m) [matrix of math nodes,row sep=2em,column sep=5em,minimum width=2em] {
     \sigma_1 & \sigma_2 \\
     \othersigma_1 & \othersigma_2 \\};
  \path[-stealth]
    (m-1-1) edge [-] node [left] {$\sigma_1 \sim \othersigma_1$} (m-2-1)
            edge node [above] {$\sigma_1 \rightarrow \sigma_2$} (m-1-2)
    (m-2-1) edge [dashed] node [below] {$\othersigma_1 \rightarrow^{*} \othersigma_2$} (m-2-2)
    (m-1-2) edge [dashed, -] node [right] {$\sigma_2 \sim \othersigma_2$} (m-2-2);
\end{tikzpicture}
\end{center}

\noindent
which, coupled with the determinacy of the target's semantics, implies refinement (see Leroy~\cite{backend} for more details).
In the commuting diagram, which shows a general forward simulation,
$\sigma$ ranges over source states and $\othersigma$ over target states.
Solid lines represent premises,
dashed lines proof obligations,
$\rightharpoonup^{*}$ represents zero-or-more steps
and $\sim$ is a \textit{matching relation} between source and target program states.

Matching relations---which describe things such as the environments, what computations are to be performed next, or the functions on the call stack, in each language---are exactly the context that provide the formal meaning of what it means for the executions to be equivalent.
Defining the matching relation is usually the most intricate part of a simulation argument.
Intuitively, matching relates equivalent points in the source and target programs' execution.
For simple passes, \eg an optimization pass whose compilation scheme replaces one instruction with another (as opposed to translating one source construct to multiple target constructs), the matching relation is easy to define, as the execution moves in lockstep.
\compiler is, of course, more complicated, further underscoring that it was a key pass to study.

\subsection{\compiler Theorem and Matching Relation}

\begin{theorem} \label{rtlgenttoplevel} \textbf{(Semantic Preservation \compiler)}\\
Let $S$ and $\mathcal{P}$ be a \source program and policy respectively. Let $C$ be the \target result of compiling $S$ and $\mathcal{I}(\mathcal{P})$ the result of compiling $\mathcal{P}$.
Under their respective semantics and policies: If $S$ does not \textbf{Go Wrong}, then the behavior displayed by $C$ is displayed by $S$.
Formally: $\forall B. \textit{safe}(S) \rightarrow C_{\mathcal{I}(\mathcal{P})} \Downarrow B \rightarrow S_{\mathcal{P}} \Downarrow B$.
\end{theorem}
\noindent
The proof of theorem \ref{rtlgenttoplevel} is mostly straightforward; we focus discussion on subtle or novel clauses of the matching relation.

The matching relation is hierarchical. It is defined on states, which in turn requires the definition of matching on functions, atoms, tag errors, call stacks, environments \etc
\noindent
Many of these constituent matchings are straightforward, because the matched structures are either very similar in both languages (e.g., call, return, and \fs states) or because they are relatively simple (\eg atoms, tag errors, or environments).
We focus our discussion on regular states, which describe computation internal to a function.
Recall that \source regular states $\sregs{F}{s}{\pct}{k}{c}{E}$ describe the current computation with a statement-under-focus, $s$, and a (local) continuation, $k$, which describes the rest of the function body.
(All \source expressions are embedded in statements.)
\target regular states $\tregs{F}{n}{\pct}{\cs}{\Rb}$ describe the current computation with a node label $n$ pointing at an instruction in the CFG (contained in the function $F$).
To preserve semantics, we just need to ensure that \source and \target functions (in regular states) step to a return state at the same time, carrying matching atoms.
However, the only way to guarantee this is to make sure that the computation in $s$ and $k$ match the computation starting at $n$, \ie that they update environment $E$ and register bank $B$ in parallel, that they make the same intermediate computations, \etc
Thus, we need to define matching for statements, continuations and expressions.
Each of these individual relations
carries pertinent information, such as the state of the register bank or specially pre-defined return registers, which we elide until we discuss particular clauses of the relation.

We match each statement $s$ to a CFG \textit{interval} [$n_e$,$n_s$), written $s\ {\sim}\ [n_e,n_s)$.
An interval is a contiguous chain of instructions (similar to the diagrams in \S\ref{compiler}) that starts at the instruction labeled $n_e$ and ends at an instruction whose successor is $n_s$. 
(The instruction at $n_s$ is not part of the interval.)
Intervals need not be \textit{linear}; they may branch so long as the last instruction in each branch has $n_s$ as a sucessor.
Similarly, $e \sim_r [n_e,n_s)$ denotes that expression $e$ matches an interval, where $r$ is the register in which
the expression's atom will be stored by the code in the interval.
Expression and statement matching are naturally closely related to their compilation schemes (cf \S\ref{compiler}).
Here are some simple cases:

\noindent
\textbf{Skip} \wskip $\ {\sim}\ [n_s,n_s)$. (Here the interval $[n_s,n_s)$ is empty.)

\noindent
\textbf{Seq} $s_1; s_2 \ {\sim}\ [n_e,n_s)$ if\ $\exists n_1$ s.t. $s_1 \ {\sim}\ [n_e,n_1) \ \land \ s_2 \ {\sim}\ [n_1,n_s)$.

\noindent
\textbf{Add} $e_1 + e_2 \sim_r [n_e,n_s)$ if $\exists n_1,n_2,r_1,r_2$ s.t.
\begin{adjustwidth}{1em}{}
\noindent $\bullet$
	$e_1 \sim_{r_1} [n_e, n_1)$ and $e_2 \sim_{r_2} [n_1, n_2)$\\

\noindent $\bullet$
	$n_2$: $\texttt{op}_+\ r_1 \ r_2\ r\ n_s$ \\
\noindent $\bullet$
	sundry side conditions, e.g.: $[n_1,n_2)$ does not overwrite $r_1$, $[n_e,n_s)$ does not overwrite registers holding variables of the \source environment, \etc
\end{adjustwidth}
\noindent
Function termination provides some more complex examples of matching relations.
In \source, a function terminates when a \wreturn\ statement is under focus, or when a function ``falls through'' by reaching the end of its body (\wskip\ under focus and empty $k$).
When falling through, functions return a default atom $a_d$.
\target functions have a single exit point, so function compilation predefines a return value register $r_{\textit{ret}}$, 
and two instructions:
\begin{adjustwidth}{1em}{}
\noindent $\bullet$
	$n_{\textit{def}}$ : $\rmovi \ a_d \ r_{\textit{ret}} \ n_{\textit{ret}}$ \\
\noindent $\bullet$
	$n_{\textit{ret}}$ : $\rret\ r_{\textit{ret}}$
\end{adjustwidth}
\noindent
The matching relation for \wreturn\ statements may look disconcerting upon first examination:
label $n_s$ is free because the \rmov jumps to the pre-defined exit $n_{\textit{ret}}$.

\noindent
\textbf{Return} \wreturn\ $e \sim [n_e,n_s)$ if $\exists n_1, r_1$ s.t.
\begin{adjustwidth}{1em}{}
\noindent $\bullet$
	$e \sim_{r_1} [n_e,n_1)$\\
\noindent $\bullet$
	$n_1$: $\rmov \ r_1 \ r_{\textit{ret}} \ n_{\textit{ret}}$
\end{adjustwidth}
\noindent
Continuation matching is defined in terms of a \textit{single} CFG node, as illustrated by the fall-through continuation.

\noindent
\textbf{Fall through} \emp\ $\sim n_{\textit{def}}$ if
\begin{adjustwidth}{1em}{}
\noindent $\bullet$
	$n_{\textit{def}}$ : $\rmovi \ a_d \ r_{\textit{ret}} \ n_{\textit{ret}}$ \\
\noindent $\bullet$
	$n_{\textit{ret}}$ : $\rret\ r_{\textit{ret}}$
\end{adjustwidth}
$n_{\textit{ret}}$ is effectively the end of the interval corresponding to every continuation.

The matching clauses we have seen so far are only slightly modified from \compcert's RTLgen. The next clause is novel, however.\\
\noindent
\textbf{k-Join} $\kjoin {\jrule} {\jpct}{k}\ {\sim}\ n_e$ if $\exists n_1$ s.t.
\begin{adjustwidth}{1em}{}
\noindent $\bullet$
	$n_e$: $\rmov \ \rpc \ \rpc \ n_1 \ \textit{@} \ \mathit{itag}$\\
\noindent $\bullet$
	$k \ {\sim}\ n_1$\\
\noindent $\bullet$
	$\forall q, {q_s} _{\in \pctags}. \ \rmovtr \ \mathit{itag} \ q \ q_s \ \_ = \jrule \ q \ q_s $\\
\noindent $\bullet$
	$\rpc$ contains $p_s$.
\end{adjustwidth}
\noindent
The first two conditions shows that an \source join rule on top of the local continuation matches to the instruction at $n_e$.
Recall that this is a pseudo-instruction that will invoke a \target level join rule (cf \S\ref{compiler}).
Consider \stepskipjoin\ (Fig. \ref{sourcestepsstmt2}): in order to process join points uniformly, it abstracts over the rule \jrule.
Without this abstraction, we would need an additional transition rule for each type of join point (there are three), in both \source and \target semantics.
As shown in the {\tt If-Then-Else} compilation diagram, split and join point instructions are generated at the same time; hence, it is easy to ensure they get corresponding split and join rule {\itag}s.
During execution, however, arbitrary code runs between the split and join points; hence, the third condition (relating \rmovtr and \jrule) insists that the \rmov carries the corresponding (and therefore correct) \itag, \ie that (\rmovtr $\mathit{itag}$) invokes \jrule.
The last condition, ensuring that \rmovtr is actually invoked on the split point PC, is really an invariant---one whose maintenance is quite involved.
We have to parameterize all compilation functions with stacks (to account for nesting of split/join points) of these ``save'' registers, and augment all matching relations with the invariant that none of the described computations trample these ``save'' registers.

To summarize matching over regular states:\\
\noindent
\textbf{Regular states}\  $\sregs{F}{s}{\pct}{k}{c}{E}\ {\sim}\ \tregs{F'}{n}{\pcp}{\cs'}{\Rb}$ if $\exists n_1$ s.t.
\begin{adjustwidth}{1em}{}
\noindent $\bullet$
	$s \sim [n,n_1) \ \land\ k \sim n_1$\\
\noindent $\bullet$
	$F \ {\sim}\ F' \ \land \ p\ {\sim}\ p' \ \land \ c\ {\sim}\ c' \ \land \ E\ {\sim}\ B$ \\
	As previously mentioned, we elide the details of these straightforward matchings
\end{adjustwidth}

\section{Optimizations} \label{optimizations}

We have analyzed the \compcert RTL-improving passes Deadcode, CSE (common sub-expression elimination), and ConstProp to 
determine what information about policies is needed to adapt these optimizations to a tagged setting. 
All these passes have the effect of removing instructions, so the key concern is whether it is valid
to skip the corresponding tag rule executions as well.
As mentioned in \S\ref{compilation}, we have identified several simple and intuitive conditions on tag rules that,
in various combinations, are sufficient to keep these optimizations sound.  These conditions are dynamic, and not
decidable in general at compile time, but there are simple conservative static approximations for each of them.
We now consider the optimizations in turn, defining the conditions as they become relevant.

\subsection{The \deadcodet pass} \label{deadcode}

{\compcert}'s Deadcode removes reachable but redundant instructions.
In ordinary RTL, an instruction is dead if its destination register is dead, and
a register is dead if it is not passed as an operand to any following instruction (before being re-defined).
In \target, since instructions and their tag rules get their operands from the same registers, the standard notion of register liveness still holds, as whenever a value is passed to an instruction, its tags are passed to that instruction's tag rule.

However the standard notion of instruction deadness is \textit{not} sufficient in \target, because even
if an instruction's result value is not used, its tag rule might still \fs, change the PC tag,
or change internal tag policy state. The latter two forms of state behave very similarly;
since our prototype compiler does not support internal tag policy state, we consider only
PC tag changes in the remainder of the paper. 

In our adaptation \deadcodet, we found that the conjunction of two conditions on rule evaluation were sufficient for a tag rule to be treated as dead:

\begin{itemize}
\item
The rule never {\fs}s. 
	\[\dfs(\ruletr) \triangleq \forall \vec{x}.\ \ruletr\ \vec{x} \not \treval \err\]
This condition is in fact \textit{necessary} to allow rule execution to be skipped: if there is a chance a rule might \fs, then skipping it might not preserve \fs behavior of the program.
\item
The rule outputs the same PC tag that it received, if it does not \fs, i.e. the rule exhibits
``PC-purity''.
\[\lpcp(\ruletr) \triangleq \forall p \ \vec{x}.\ \ruletr\ p\ \vec{x} \treval (p',\ \vec{x'}) \rightarrow p = p'\]
Since the PC tag is threaded throughout the program's execution, it can effectively be used to pass state, which could affect a \fs decision in a later rule.
    \lpcp simply says that a rule is side-effect-free with respect to the PC tag.

\end{itemize}

\noindent
($\dfs \land \lpcp$) is used as an additional guard to the standard notion of instruction deadness, both in the liveness analysis and the code transformation.
Although these conditions are not statically computable, they have simple conservative approximations:
A rule must be \dfs if it never returns a tag error, and must be \lpcp if the output PC tag is always syntactically equal to the input PC tag.

\subsection{\deadcodet Theorem and Verification}

\begin{theorem} \label{deadcodetoplevel} \textbf{(Semantic Preservation \deadcodet)}
Let $S$ and $\mathcal{P}$ be a \target program and policy respectively, and $C$ be the \target result of performing the deadcode optimization on $S$.
Under \target semantics and the policy $\mathcal{P}$: If $S$ does not \textbf{Go Wrong}, then the behavior displayed by $C$ is displayed by $S$.
Formally:
$\forall B. \textit{safe}(S) \rightarrow C_{\mathcal{P}} \Downarrow B \rightarrow S_{\mathcal{P}} \Downarrow B$.
\end{theorem}
\noindent
Having covered the generalities in \S\ref{vprelim}, we discuss here how the analysis and proof of theorem \ref {deadcodetoplevel} are driven by a set of \textit{flags} that indicate which properties hold on the \source rules.
As optimizations work over \target, it is the properties of \target rules that we are interested in.
Just as we define \target policies out of \source ones, we define \target flags out of \source ones and prove that whenever a \target rule has a property, so does the corresponding \source rule. \eg \lpcp(\rmovtr \itass) only if \lpcp(\wassigntr).

In the correctness proof, we would like these flags to have type \texttt{Prop}, but in the compiler, we need them to be computable.
So, we encode them as \texttt{option} \texttt{Prop}s.
For example, \texttt{Some} ($H$:\dfs(\roptr)) tells the compiler that \roptr does not \fs and
provides the \texttt{Prop} $H$:\dfs(\roptr) for use in the proof.
The \texttt{None} case tells the compiler the property does not hold.

This dependent type does double duty: It helps us cleanly define \target flags out of \source ones while simultaneously verifing that if a \target rule has property $X$, so does its related \source rule. 
We define the \target flags in Coq proof mode by providing a \texttt{Some} (requiring a proof of the property it carries) or \texttt{None} witness.
Eliminating cases on \source flags in order to provide such witnesses defines \target flags from \source ones.
This approach helps us \textit{validate} our policy compilation.
As an example, if we wanted to show a witness for \texttt{Some} ($H$:\dfs(\roptr)), not having to case analyze the \source flag (\texttt{option} \dfs($\oprule$)) would be a hint that the definition of \roptr (via compilation of the \source policy) is wrong.
This validation mechanism caught several bugs in our initial compiler code.

There is a caveat attached to our verification.
In \tagine's current implementation, we only model \source policy signatures, not actual rule definitions.
Therefore, the compiler cannot derive the \source flag settings by inspecting the policy, but instead relies on external
specification of the flag settings as axioms.
As future work, we envision modelling the policy rule language in detail, so that properties of \source policies can be extracted
by a provably-correct static analysis.

\subsection{\cse (Common-Subexpression Elimination)} \label{cse}

This pass replaces repetitions of an \rop instruction (the common sub-expression) with a \rmov instruction that writes the previously computed value into the \rop destination register.

The variety of CSE implemented by \compcert is local value numbering (LVN).
LVN works by maintaining a bijective mapping between symbolic identifiers (the value numbers) and expressions (\ie variables or operations).
It operates as a forward dataflow analysis over extended basic blocks.
When encountering a (syntactic) expression, LVN checks the map to see if it already has a value number; if not, a fresh value number is assigned to the expression, and the map is updated accordingly. 
In standard LVN, an expression's value number is cleared if any of its constituent variables are redefined.
\setlength{\multicolsep}{2pt}
    \begin{multicols}{2}
    \noindent
    \begin{center}
    \begin{minipage}{0.5\linewidth}
    \verb|1: z = x + y|\\
    \verb|2: c = a + b|\\
    \verb|3: w = x + y|\\
    \verb|4: x = 5|\\
    \verb|5: v = x + y|
    \begin{center}
    pre-CSE
    \end{center}
    \end{minipage}
    \end{center}

    \columnbreak
    
    \noindent
    \begin{center}
    \begin{minipage}{0.5\linewidth}
    \verb|1: z = x + y|\\
    \verb|2: c = a + b|\\
    \verb|3: w = z|\\
    \verb|4: x = 5|\\
    \verb|5: v = x + y|
    \begin{center}
    post-CSE
    \end{center}
    \end{minipage}
    \end{center}

    \end{multicols}
\noindent
In the pre-CSE pseudocode, while lines 1, 3 and 5 contain a syntactically equivalent expression (\texttt{x + y}), only lines 1 and 3 have a \textit{common} sub-expression, as they perform the same computation, while line 5 does not, due to the redefinition of \texttt{x} on line 4.
LVN determines this by assigning the first repetition (line 3) the same value number as the original (line 1), because nothing causes it to be cleared from the map.
However, the repetition on line 5 gets a new value number because \texttt{x} is redefined on line 4.
LVN then replaces repeated sub-expressions whose value number is associated with a variable by an \rmov from that variable, as illustrated in the post-CSE pseudocode.

The standard notion of LVN guarantees that two expressions with the same value number are equivalent computations.
In the \tagine setting, this is enough to guarantee that the rules of two expressions with the same value number will receive the same value tag inputs,
but we still need to account for the PC tag input. 
The intuition is that LVN is sound in \tagine whenever \roptr is insensitive to the PC tag input or the repeated rules receive the same PC tag input as the original.
We present two cases where this holds.

\textbf{(a)}
When the \texttt{op}'s rule is (weakly) insensitive to the PC tag input, meaning that it is \lpcp, and its PC tag input influences neither its output value tags nor whether it {\fs}s.

\begin{center}
\begin{tabular}{rl}
    \multicolumn{2}{l}{$\wpci(\textit{rule}) \triangleq \lpcp(\textit{rule}) \land (\forall p_1,\ p_2, \ \vec{\v}.$}\\
    $(\forall p_1'\ p_2'\ \vec{\v_1'}\ \vec{\v_2'}.$ & $\ruletr\ p_1\ \vec{\v} \treval (p_1', \vec{\v_1'}) \ \land$ \\
    & $\ruletr\ p_2\ \vec{\v} \treval (p_2', \vec{\v_2'}) \rightarrow \vec{\v_1'} = \vec{\v_2'})$ \\
    \multicolumn{2}{l}{$\land\ \ruletr\ p_1\ \vec{\v} \treval \err \leftrightarrow \ruletr\ p_2 \vec{\v} \treval \err$)}
\end{tabular}
\end{center}

\noindent
The intuition for PC insensitivity is that a rule should ``do nothing'' with the PC tag, and in the case of weak PC insensitivity, propagating the input PC tag is the most innocuous choice for the output PC tag.
In this case, the standard definition of LVN is already sound in \tagine.

\textbf{(b)}
When a repeated \texttt{op}'s rule can be guaranteed to receive the same PC tag input as the original because all intervening instructions between the original sub-expression (including that sub-expression itself) and a candidate repetition are \lpcp.

In summary, LVN in \tagine must modify standard LVN to clear out the value numbers of all non-\wpci instructions upon encountering a non-\lpcp instruction.

We have implemented this revised version of \cse in \tagine, but have not completed its verification.

\subsection{ConstProp} \label{constprop}
\cp folds constants (concrete values known at compile time) by turning \ropns s whose results can be computed at compile time into {\rmovi}s.
It also performs constant propagation by a dataflow analysis over the contents of registers to compute their  abstract values.

As a running example, consider a register bank with $ r_1 \deff  3,  r_2 \deff 4$.
Standard constant folding makes the change:
\begin{center}
    $\texttt{op}_+ \ r_1\ r_2\ r_d \Rightarrow \rmovi\ 7\ r_d$
\end{center}
\noindent
In \tagine we need to compute a tag to write into $r_d$ as well.
We outline two approaches to making folding sound in \tagine.

The first approach applies when the \rop to be folded has constant operand tags.
This approach permits folding by (ultimately) invoking \roptr despite replacing the \rop with a \rmovi.
It does so with a special \itag that can take parameters, \eg (\texttt{ITp} $\oplus\ t_1\ t_2$),
where everything enclosed by the parentheses is one I-tag.
In our running example, with $r_1 \deff 3 \textit{@} c_1, r_2 \deff 4 \textit{@} c_2$, folding makes this change:
\[
\texttt{op}_+ \ r_1\ r_2\ r_d \ \textit{@} \ \itop + \Rightarrow \rmovi\ (7 \textit{@} \_) \ r_d\ \textit{@}\ (\texttt{ITp} +\ c_1\ c_2)
\]
\rmovitr is defined to invoke $\overline{\texttt{op}_+}$ on $c_1$ and $c_2$ when given {\itag} $(\texttt{ITp} +\ c_1\ c_2)$, and $r_d$ is tagged with the result.

The second approach applies when we can statically compute a concrete output value tag for {\roptr}, which occurs in the following cases:
\textbf{(1)} If \roptr is \lpcp
\footnote{In \source, expression rules are designed to be implicitly \lpcp: they do not even produce an output PC tag. 
Therefore \target rules defined from \source expression rules will always be \lpcp.}
and produces a constant value tag, implying insensitivity to all its inputs. In this case we do not require concrete inputs.
\textbf{(2)} If \roptr is \lpcp, its value tag output is simply propagated from one of its inputs, and that input is known at compile time.
\textbf{(3)}
If \roptr is strongly PC insensitive (\ie the PC tag input does not influence the rule's output), the input value tags are known, and we can evaluate the rule at compile time.
If the computed result is a \fs, the pass does not replace the \rop, preserving \fs behavior.

Neither of these approaches is strictly more useful or applicable than the other. 
The first approach requires concrete tag values but can deal with dynamic PC tag input.
This approach however, also generates more I-tags, which will cause more compulsory misses.
The second approach does not always require concrete tag values, but is only applicable in rather ad hoc circumstances.

\section{Coq Development}\label{coqdev}

The proof of \compiler is ${\sim}1700$ lines, while \deadcodet has a proof of ${\sim}650$ lines.
These numbers reflect formalization specific to the passes.
From \compcert, we also used some proofs on the general metatheory of simulations.

The goal of keeping the components of \tagine as decoupled as possible led us to adopt a highly modularized and functorized architecture in Coq.
In particular, IRs and compiler passes do not depend on semantics or proofs.
However, to implement monitoring at \target-level, \tagine must invent new \target tags and policies.
Moreover, while the abstract definition of \target is independent of \source, all \tagine-\target notions (tags, language, policies, flags, semantics) must be parameterized by \source ones.
This means that optimization passes must be functors over \tagine-\target and therefore parameterized by \source tags, policies and flags as well.
As examples, the proofs of \compiler and \deadcodet are functors over eleven and eight other modules, respectively.

\section{Related work} \label{related}

Hardware reference monitors and other secure hardware platforms have been the focus of
much recent attention as a potential foundation for secure systems. Some target a specific
security policy; for example, CHERI~\cite{CHERI} implements compartmentalization using
capabilities. PIPE instead aims to be general~\cite{PUMP-ASPLOS,PUMP-HASP}.
It has been used to enforce information flow control~\cite{PIPE-IFC},
stack safety~\cite{stack}, and capability-based heap-safety, among other
micropolicies~\cite{micropol}. Abate et. al. use PIPE as an example enforcement mechanism
for their Secure Compartmentalized Compilation property \cite{goodbad}.

Aspect-Oriented Programming bears a structural similarity to the reference monitor
approach; when used for security it also entails interleaving policy validation
with application code~\cite{KiczalesLMMLLI97}. Advice points are akin to our control points.
But AOP's advice code is normally written in the same language as the underlying
program and can operate on the full program environment,
which naturally suggests a semantics and implementation based on weaving together
the program and advice. 
While parts of AOP semantics have been formalized~\cite{DouenceMS01,aspects},
we are not aware of any attempts to prove correctness of AOP tool implementations.

Like our work, much of the literature in compiler verification focuses on toy compilers that illustrate a key challenge~\cite{compCorrMech}, or verifies compilation of a specific, small part of a language~\cite{arithExprCorr}.
The VLISP~\cite{vlisp} project is notable in that it has a correctness proof for an implementation of LISP. But while rigorous, it is not machine checked.
\compcert~\cite{backend} and \cakeml~\cite{cake} stand alone as industrial strength, machine-checked, verified compilers;
the former has been used to explore verifying optimizations, while \cakeml has focused on reducing the trusted computing base, and verifies other parts of the run time, such as the garbage collector.
\balance

\section{Conclusions and Future Work} \label{future}

We have demonstrated a plausible design for high-level tag-based monitoring and its compilation to \pipe-equipped hardware,
formalized a prototype compiler,
and verified that it preserves monitoring semantics.
Although our formal development covers only a toy source language, it has allowed us to confirm the feasibility of the most novel aspects of the compilation approach.

There are numerous ways to extend this work to handle more realistic source languages and compilation mechanisms,
in particular towards our goal of a fully verified compiler for a tagged version of C.
Our first priority is to add addressable memory and pointers. 
In particular, we are interested in using policies to enforce memory safety and compartmentalization properties
on top of a memory-unsafe but control-safe language called Concrete C, with the goal of giving security
engineers a flexible tool for trading different levels of memory safety against performance. 
Adding support for this feature should be largely orthogonal to the existing \tagine development, although
we will need to extend the optimization passes to handle pointer operations.

We also plan to extend \source and \compiler to handle the full set of C control flow operators. We expect
this task to be straightforward, although we are still exploring whether useful IFC policies can be defined
for this richer language.

\begin{acks}
We thank the anonymous reviewers for their valuable feedback. 
Greg Sullivan and C\u{a}t\u{a}lin Hri\cb{t}cu provided useful comments on earlier drafts.
This material is based upon work supported by the Defense Advanced Research Projects Agency (DARPA) under Contract No. HR0011-18-C-0011.
Any opinions, findings and conclusions or recommendations expressed in this material are those of the author(s) and do not necessarily reflect the views of the Defense Advanced Research Projects Agency (DARPA).
\end{acks}

\bibliography{biblio}

\end{document}